\documentclass[prb,twocolumn,showpacs,superscriptaddress,amsmath,floatfix]{revtex4}
\usepackage{graphicx}
\usepackage{amssymb}
\usepackage{color}
\usepackage{ulem}

\begin{document}

\title{Theory of metal-intercalated phenacenes: Why molecular valence 3 is special}
\author{Tirthankar Dutta} 
\author{Sumit Mazumdar}
\affiliation{Department of Physics, University of Arizona, Tucson, AZ 85721} \date{\today}
\begin{abstract}
We develop a correlated-electron minimal model for the normal state of charged phenanthrene ions in 
the solid state, within the reduced space of the two lowest antibonding molecular orbitals of 
phenanthrene. Our model is general and can be easily extended to study the normal states of other 
polycyclic aromatic hydrocarbon superconductors. The main difference between our approach and 
previous correlated-electron theories of phenacenes is that 
our calculations are exact within the reduced basis space, albeit for finite clusters.
The enhanced exchange of electron 
populations between these molecular orbitals, driven by Coulomb interactions over and above
the bandwidth effects, gives a theoretical description of the 
phenanthrene trianions that is very different from previous predictions. Exact many-body finite 
cluster calculations show that while the systems with molecular charges of $-$1 and $-$2 are one- and 
two-band Mott-Hubbard semiconductors, respectively, molecular charge $-$3 gives two nearly 
$\frac{3}{4}$-filled bands, rather than a completely filled lower band and a $\frac{1}{2}$-filled 
upper band. The carrier density per active molecular orbital is thus nearly the same in the normal 
state of the superconducting aromatics and organic charge-transfer solids, and may be the key to 
understanding unconventional superconductivity in these molecular superconductors.
\end{abstract}
\maketitle

\section{Introduction}
\label{intro}
Superconducting metal-intercalated polycyclic aromatic hydrocarbons (PAHs) 
\cite{Mitsuhashi10a,Kubozono11a,Wang11a,Xue12a}, discovered recently, constitute the third family of 
molecular carbon(C)-based superconductors, besides  organic charge-transfer solids (CTS) 
\cite{Ishiguro} and fullerides \cite{Gunnarsson97a}. Superconductivity (SC) has been detected from 
magnetic measurements in alkali metal (K, Rb)$-$intercalated picene \cite{Mitsuhashi10a}, 
coronene \cite{Kubozono11a}, phenanthrene \cite{Wang11a} and dibenzopentacene (T$_c$ $>$ 30 K) 
\cite{Xue12a}. Zero resistance has been confirmed in K$_3$-picene \cite{Teranishi13a}. Experiments 
in all cases reveal two rather remarkable observations, viz., (i) SC occurs in phenacenes 
(phenanthrene and picene) with ``armchair'' edges and PAHs with related structural motif, while it is 
absent in metal-intercalated acenes with linearly fused benzene rings, and, (ii) ``doping'' of nearly 
3 electrons per PAH molecule is essential for SC. For example, in K-doped 
picene, hereafter K$_x$picene, SC has been found both at ``low T$_c$'' of 7 K, 
and ``high T$_c$'' of 18 K in annealed samples. However, in samples produced from solutions only the 
18 K superconductor is obtained with $x$ lying within a narrow range of $2.9-3.1$ \cite{Kubozono11a}. 
Spectral shift measurements of molecular Raman modes have also indicated that the charge on the 
picene molecules at the superconducting compositions is nearly exactly $-3$, independent 
of the nature of the metal ions. Very similar behavior is found in the metal-intercalated 
phenanthrenes \cite{Wang11a}. Observations of SC in nearly stoichiometric Ca$_{1.5}$picene 
\cite{Kubozono11a}, Sr$_{1.5}$ and Ba$_{1.5}$ phenanthrene \cite{Wang11b}, and, La- and 
Sm-phenanthrene \cite{Wang12a} have further confirmed the limitation of SC to ionic compounds with 
charges of $-3$ on the PAH molecules. Raman spectral shifts in 
the doped phenanthrenes are independent of the metal ions, and again correspond to molecular
charge $-3$. Interestingly, the strong peaking of T$_c$ at this particular doping is also shared by 
the fullerides \cite{Gunnarsson97a}.

The above observations place severe but obvious constraints on the correct theory of the normal 
states of charged phenacenes. Development of such a correct theory is the 
crucial first step to the future development of a theory of SC in these complex materials. There 
exist currently no theory of the normal state of the doped PAHs that 
explains both observations (i) and (ii) noted above. Doping of upto 3 electrons per molecule leads to 
occupancies of both the lowest unoccupied molecular orbital LUMO (L) and the next higher level LUMO+1 
(L+1) by the doped electrons. A significant difference between acenes on the one hand, and phenacenes 
and coronene on the other, is that the single-particle energy gap between L and L+1, $\Delta_{L,L+1}$,
is much smaller in the second group of molecules (see Fig. 1 and Table I below), 
which immediately suggests that small $\Delta_{L,L+1}$ is essential for 
SC. Indeed, density 
functional theory (DFT)-based band calculations for K$_x$picene 
\cite{Kosugi09a,Kim11a,deAndres11a,Kosugi11a} and K$_3$phenanthrene \cite{deAndres11b} find 
significant hybridization between the bands derived from the L and L+1 MOs; however, 
DFT calculations generally underestimate both $\Delta_{L,L+1}$ and the gap 
$\Delta_{HL}$ between the highest occupied molecular orbital, HOMO, and LUMO 
\cite{Kubozono11a}.

DFT-based band calculations have led to strong coupling 
BCS-based multi-band theories of SC for doped picene that emphasize band hybridization and, 
intermolecular \cite{Casula12a} and intramolecular \cite{Kato11a,Subedi11a,Casula12a} 
electron-lattice vibration couplings. These theoretical formulations however do not give 
natural explanations for the absence of SC in the doped acenes; even in the absence of
band overlap in these latter systems the same electron-phonon couplings should have driven SC,
albeit at lower T$_c$.
Additionally, there is nothing unique about anions with charge $-3$ within these theories.
The observation of dT$_c$/dP $>$ 0 (where P = pressure) in high $T_c$ K$_x$picene \cite{Kambe12a}, 
and in the superconducting phenanthrenes, with pressure coefficient nearly independent of the cation 
in the latter \cite{Wang11a,Wang11b,Wang12a}, also argues against the BCS approach.

Taken together, the above observations have led to the belief that PAH superconductors are 
unconventional, and SC may be driven by repulsive electron-electron (e-e) interactions. The 
interacting electron picture has received support from the observation of Mott-Hubbard semiconducting 
behavior in K$_1$pentacene \cite{Craciun09a}, in which $\Delta_{L,L+1}$ is much larger than in the 
phenacenes (see Fig.~1 and Table I). Interacting electron theories that have been constructed for the doped acenes 
and phenacenes \cite{Giovannetti11a,Huang12a,Ruff13a}, however, still fail to explain the two crucial 
experimental observations completely. Taking the combined effects of the Hubbard $U$ and nonzero 
$\Delta_{L,L+1}$ into consideration, reference \onlinecite{Giovannetti11a} has concluded that 
K$_3$picene in the normal state is a Mott-Hubbard semiconductor, with a completely filled L-band and 
a $\frac{1}{2}$-filled L+1-band. Very similar description is also obtained from calculations based on 
combined DFT and dynamical mean-field theory (DMFT), which have found antiferromagnetic 
semiconducting behavior for K$_x$picene for all integer $x$ \cite{Ruff13a}. These theories do not 
provide the correct starting point for any theory of SC in the trianions. Although mean field and 
DMFT calculations had found antiferromagnetic-to-SC transition within the frustrated 
$\frac{1}{2}$-filled Hubbard band model in the past \cite{Schmalian98a,Vojta99a,Kyung06a,Powell},
more recent numerically precise calculations, using a variety of techniques, have universally found 
absence of SC within this model 
\cite{Clay08a,Dayal12a,Tocchio09a,Watanabe08a,Gomes13a,Yanagisawa13a}. 
From an experimental perspective, the absence of SC in  the Mott-Hubbard semiconductor K$_1$pentacene 
\cite{Craciun09a} confirms the theories that predict absence of SC within the $\frac{1}{2}$-filled 
band Hubbard model. $\Delta_{L,L+1}$ approaching zero in picene, as has been recently suggested 
\cite{Kim13a}, does not provide a solution either, both because such tiny $\Delta_{L,L+1}$ is 
improbable (see below), and also because this once again fails to explain the specific role of 
molecular charge of $-3$. 

To summarize, we are faced with a  conundrum. Electron-phonon coupled models predict metallic 
behavior for all molecular charges, including $-2$ which is known to exist experimentally, and 
therefore cannot explain the restriction of SC to molecular charge $-3$. Models incorporating e-e 
interactions predict semiconducting behavior at all integer charges. There is thus an obvious need for 
the development of a correct theory of the {\it normal state} of the doped phenacenes and coronene, 
which can at least provide the starting point for a plausible theory of correlated-electron SC in 
these systems. In the present paper, we develop a correlated-electron minimal model for lattices of 
PAH anions that can explain both the crucial observations we have made in the above. Our approach is 
general and applicable to both phenacenes and acenes (see Appendix). For the specific case of 
phenanthrene ions we show from exact many-body cluster calculations that while crystals with 
molecular charges of $-1$ and $-2$ are indeed one- and two-band Mott-Hubbard semiconductors, in 
agreement with previous work \cite{Ruff13a}, crystals of trianions consist of two nearly 
$\frac{3}{4}$-filled bands, even with realistic $\Delta_{L,L+1} \geq 0.2$ eV. This specific 
band-filling is also characteristic of superconducting CTS \cite{Ishiguro}, as well as, a  variety of 
apparently unrelated inorganic superconductors \cite{Mazumdar12a,Mazumdar14a}. One of us and 
colleagues have shown that precisely at this band-filling there is a strong tendency to form a 
correlated-electron spin singlet-paired semiconductor \cite{Li10a,Dayal11a}, which may undergo 
semiconductor-superconductor transition \cite{Mazumdar08a}. While the theory of 
semiconductor-superconductor transition within the model constitutes ongoing research, examples of 
the spin-paired semiconductor already exist and our present results give additional credence to the 
theory.

In section II we present our theoretical model of PAH anions, starting from an {\it atomic basis} of 
$\pi$ electrons on the carbon atoms of the molecules. We include the Hubbard repulsion between the 
$\pi$-electrons at the outset, and thus our approach is different from theories that ``graft on'' e-e 
interactions to band theoretical results. Starting from this fundamental atomic basis we derive an 
exact effective Hamiltonian in the reduced space of L and L+1 MOs. 
The effective Hamiltonian contains a term involving two-electron hops  
that has been ignored in previous correlated-electron models of phenacenes 
\cite{Giovannetti11a,Huang12a,Ruff13a,Kim13a}, but {\it should not be ignored for small 
$\Delta_{L,L+1}$}, for the sake of completeness. However, as we demonstrate later, the effects of
this off-diagonal Coulomb interaction on MO populations is weak compared to that due to 
diagonal density-density Coulomb terms.
As is shown numerically in section III, the two-band 
$\frac{3}{4}$-filled nature of the trianion system is a consequence of {\it co-operation} between 
bandwidth and correlation-induced effects. 
Physical, intuitive reasonings that explain the {\it mechanism} of the electron population exchange between
the MOs, and detailed understanding of the numerical results, are presented in the Appendix.
In section IV, we summarize our conclusions and discuss a 
plausible (albeit incomplete) theory of correlated-electron SC in the PAH superconductors in the 
context of previous work \cite{Li10a,Dayal11a,Mazumdar08a}. We also briefly discuss the 
antiferromagnetic-to-SC transition in Cs$_3$C$_{60}$ \cite{Ganin08a,Ganin10a,Capone09a} from the 
perspective of the present results.  

\section{Theoretical model: Complete and reduced basis spaces.}

We derive and study a model Hamiltonian describing the normal state of lattices of PAH 
or acene anions. The dopant metal ions are not included explicitly in the Hamiltonian. 
Justification for ignoring the metal ions comes from the nearly universal behavior of the intercalated 
phenanthrenes and picenes, independent of the metal ions 
\cite{Wang11a,Wang11b,Wang12a,Mitsuhashi10a,Kubozono11a}. In particular, antiferromagnetic 
ordering of Sm$^{3+}$ ions in Sm-phenanthrene at T$_N \sim15$ K, and the coexistence of SC with this 
antiferromagnetic ordering \cite{Wang12a} indicates the weak role of the metal ions.
In agreement with the DFT calculations \cite{Kosugi09a,Kim11a,deAndres11a,deAndres11b,Kosugi11a} we 
assume that there occurs a homogeneous distribution of the ions in the actual materials, and in 
addition to donating electrons, they also enhance the intermolecular hoppings relative 
to the pristine systems by creating chemical pressure. We write the minimal many-body Hamiltonian for 
the organic ions in the solid state as
\begin{equation}
H=H_{intra}^{1e}+H_{intra}^{ee}+H_{inter}^{1e}
\label{ham}
\end{equation}
where $H_{intra}^{1e}$ and $H_{intra}^{ee}$ are the one-electron and many-electron components of the intramolecular Hamiltonian,
and $H_{inter}^{1e}$ is the intermolecular hopping.
The individual terms are written as,
\begin{eqnarray}
H_{intra}^{1e}&=&-\epsilon \sum_{\mu,i}{}^{'} n_{\mu,i} - t\sum_{\mu,\langle ij \rangle,\sigma}c_{\mu,i,\sigma}^{\dagger}c_{\mu,j,\sigma} \\
H_{intra}^{ee}&=&U\sum_{\mu,i,}n_{\mu,i,\uparrow}n_{\mu,i,\downarrow} \\
H_{inter}&=&\sum_{\mu \neq \nu,i,j,\sigma}t_{\mu,\nu,i,j}c_{\mu,i,\sigma}^{\dagger}c_{\nu,j,\sigma}
\label{terms}
\end{eqnarray}
In the above $c_{\mu,i,\sigma}^{\dagger}$ creates a $\pi$-electron of spin $\sigma$ in the $p_z$ 
orbital of the $i$-th {\it C-atom} of the $\mu$-th molecular ion, 
$n_{\mu,i,\sigma}=c_{\mu,i,\sigma}^{\dagger}c_{\mu,i,\sigma}$, and 
$n_{\mu,i}=\sum_{\sigma}n_{\mu,i,\sigma}$;
$t$ and $U$ are the nearest neighbor {\it intra}molecular hopping integral and effective repulsion 
between two electrons occupying the same $p_z$ orbital, respectively, and $t_{\mu,\nu,i,j}$ is the 
{\it inter}molecular hopping between C-atoms $i$ and $j$ of molecules $\mu$ and $\nu$. We have not 
included the interatomic Coulomb repulsions in $H_{intra}^{ee}$ and $H_{intra}^{ee}$. This does not lead
to loss of generality.
The effects due to the {\it intra}molecular interatomic interactions can be 
included in the effective onsite repulsion $U$, while the {\it inter}molecular interatomic 
interactions play a negligible role in what follows and can be incorporated at a later stage in the 
search for a  theory of SC if so desired, without loss of generality. The primed sum in the 
site-energy dependent term in Eq.~{2} is restricted to C-atoms without C-H bonds, accounting for 
their larger electronegativity relative to the other C-atoms \cite{Huang12a}.

The atomic basis space is complete, but the Hamiltonian is clearly unsolvable because of its very 
large dimension. We rewrite the Hamiltonian in the
MO basis, from which we then derive a reduced Hamiltonian in the space of L and L+1 MOs 
of the phenanacene ions. In this we use the standard approach of molecular exciton theory, and 
ignore one-electron as well as many-electron matrix elements involving widely separated MOs. Given 
the relatively small $\Delta_{L,L+1}$, however, Coulomb matrix elements involving two-electron hops 
between L and L+1 cannot be ignored, and this is where we differ from previous correlated-electron 
models of phenacenes \cite{Giovannetti11a,Ruff13a}. 

\begin{figure}[tb]
\includegraphics[height=2.0in,width=3.5in]{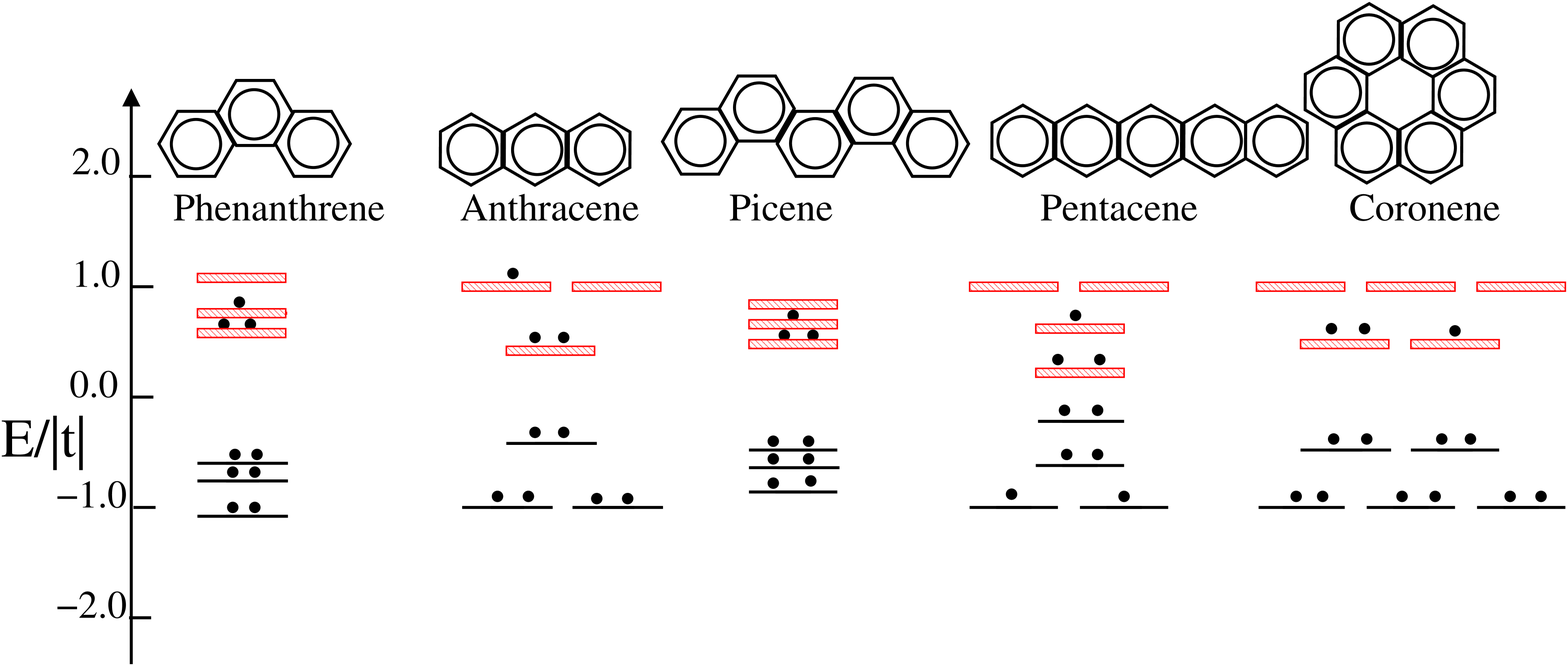}
\caption{(Color online) Bonding (thin black lines at negative energies) and antibonding (thick red lines at positive energies) MOs, 
within nearest neighbor $\pi$-electron 
only tight-binding theory and for identical
carbon atoms, near the chemical
potential of phenacenes, acenes and coronene. In all cases occupancies of the antibonding MOs by 3 extra electrons are shown, within
a rigid bond approximation. The one-electron energies are in units of $|t|$.}
\label{picene1}
\end{figure}

The first step in constructing the effective Hamiltonian in the MO basis is to solve $H_{intra}^{1e}$ exactly,
\begin{equation}
H_{intra}^{1e}=\sum_{\mu,k,\sigma}E_{k}a_{\mu,k,\sigma}^\dagger a_{\mu,k,\sigma} 
\label{Huckel}
\end{equation}
Here
$a_{\mu,k,\sigma}^{\dagger}=\sum_{i}\psi_{\mu,k,i}c_{\mu,i,\sigma}^{\dagger}$ corresponds to the
$k$th MO of the $\mu$th molecule. In Fig.~\ref{picene1} 
we have shown the highest few bonding and lowest few antibonding MOs
that are obtained as solutions to $H_{intra}^{1e}$ with $\epsilon=0$ and the same $t$ for all the molecules included in the Figure.
As mentioned above $\Delta_{L,L+1}$ in phenanthrene and picene are much smaller than those in anthracene and pentacene, respectively.
$\Delta_{L,L+1}=0$ in coronene, but is expected to be small but nonzero if the Jahn-Teller effect due to interactions of electrons
with molecular vibrations are included. Correspondingly, $\Delta_{HL}$ in the phenacenes and coronene are much larger than in
the acenes. Table I gives the numerical values of these gaps, along with the gap $\Delta_{L+1,L+2}$ between LUMO+1 and LUMO+2
in units of $|t|$, which we have taken to be 2.4 eV \cite{Baeriswyl92a}. Note however that $\Delta_{L,L+1}$ in phenanthrene (picene) is about
25\% (50\%) of that in anthracene (pentacene). Thus recent calculations that find tiny $\Delta_{L,L+1}$ (0.04 eV) for picene along 
with very large $\Delta_{L,L+1}$ for pentacene (1.26 eV) \cite{Kim13a} are unrealistic, and cannot be obtained within $H_{intra}^{1e}$ without
assuming widely different $|t|$ for these molecules.
\begin{table}[!htbp]
\caption{One-electron energy gaps between HOMO--LUMO, LUMO--LUMO+1 and LUMO+1--LUMO+2 in units of $|t|$ in the 
phenacenes, acenes, and coronene, for $\epsilon$ = 0}
\begin{tabular}{|c||c|c|c|}
\hline
Molecule & $\Delta_{H,L}$ & $\Delta_{L,L+1}$ & $\Delta_{L+1,L+2}$ \\
\hline
Phenanthracene & 1.210 & 0.164 & 0.373 \\
\hline
Anthracene & 0.828 & 0.586 & 0.000 \\
\hline
Picene & 1.004 & 0.178 & 0.179 \\
\hline
Pentacene & 0.439 & 0.398 & 0.382 \\
\hline
Coronene & 1.078 & 0.000 & 0.461 \\
\hline
\end{tabular}
\end{table}

$H_{intra}^{ee}$ and $H_{inter}^{1e}$ are now expressed in terms of
these {\it localized} MOs $a_{\mu,k,\sigma}^\dagger$ \cite{Chandross99a},
\begin{equation}
\begin{split}
H_{intra}^{ee}=U \biggl[ \sum_{\mu,k,k^\prime,i} &|\chi_{\mu,i,k}|^2|\chi_{\mu,i,k^{\prime}}|^2 N_{\mu,k,\uparrow} N_{\mu,k^\prime,\downarrow} + \\
\sum_{\mu, k_1 \neq k_2, k_3 \neq k_4,i}(\prod_{l=1}^{4}\chi_{\mu,i,k_l}) &a_{\mu,k_1,\uparrow}^{\dagger} a_{\mu,k_2,\uparrow} a_{\mu,k_3,\downarrow}^{\dagger} a_{\mu,k_4,\downarrow} \biggr] \\ 
\end{split}
\label{reduced1}
\end{equation}
\begin{equation}
H_{inter}^{1e}  = \sum_{\mu \neq \nu, k_1, k_2, \sigma} \sum_{i,j} \chi_{\mu,i,k_1} \chi_{\nu,j,k_2} t_{\mu,\nu,i,j} a_{\mu,k_1,\sigma}^{\dagger} a_{\nu,k_2,\sigma} 
\label{reduced2}
\end{equation}

Here the $\chi$ and $\psi$ matrices are inverses of one another,
$N_{\mu,k,\sigma}=a_{\mu,k,\sigma}^{\dagger}a_{\mu,k,\sigma}$, and we define $t_{\mu,\nu}^{k_1,k_2}$
$=$ $\sum_{i,j}\chi_{\mu,i,k_{1}}\chi_{\nu,j,k_{2}}t_{\mu,\nu,i,j}$.

The above transformation is exact.
We now note that the intermolecular hoppings are tiny relative to the HOMO-LUMO gaps of both acenes 
and phenacenes, and do not affect the occupations of the bonding MOs or the high energy antibonding 
MOs. The reduced Hamiltonian $H_{L,L+1}$
is therefore obtained by restricting the sums over the {\it k's} in 
Eq.~(6) and (7) to L and L+1, with $E_{L+1}-E_L=\Delta_{L,L+1}$. 
In what follows, we will distinguish between the diagonal density-density dependent terms of $H_{intra}^{ee}$, which we write
as $H^{ee}_d$, and the off-diagonal terms involving two-electron hops between MOs, which we write as $H^{ee}_{od}$.
$H^{ee}_d$ consists of three distinct terms, $U_{L,L}$, $U_{L,L+1}$, and $U_{L+1,L+1}$, given 
by,
\begin{eqnarray}
U_{L,L} &= \sum_{i}|\chi_{\mu,i,L}|^2|\chi_{\mu,i,L}|^2 \\
U_{L+1,L+1} &= \sum_{i}|\chi_{\mu,i,L+1}|^2|\chi_{\mu,i,L+1}|^2 \\
U_{L,L+1} &= \sum_{i}|\chi_{\mu,i,L}|^2|\chi_{\mu,i,L+1}|^2 
\end{eqnarray}
where $U_{L,L}$ and $U_{L+1,L+1}$ are the repulsions between two electrons occupying
L and L+1, respectively, and $U_{L,L+1}$ is the repulsion between electrons of opposite 
spins occupying different MOs of the same molecule. 
As shown in Fig.~\ref{structure}(d), $H^{ee}_{od}$ consists of two kinds of terms, both proportional to $U_{L,L+1}$.

\section{Computational Approach and Results.}

While our theoretical formulation in section II is general, we have performed numerical calculations 
explicitly for phenanthrene and anthracene ions only, as the computations quickly become 
unmanageably large and complex for the larger molecules. We discuss picene and 
coronene in the Appendix. In Fig.~\ref{structure}(b) we show the simplified two-dimensional (2D) 
herringbone structure of the doped phenanthrene crystal \cite{Wang11a} we consider. We retain 
intermolecular hoppings $t_{\mu,\nu}^{k_1,k_2}=t_{|\mu-\nu|}^{k_1,k_2}$ for $|\mu-\nu|=j=1,2$ and 
$k_1,k_2=L,L+1$, labeled $t_1$ and $t_2$ in Fig.~\ref{structure}(b). The terms $t_j^{L+1,L}$
in Fig.~\ref{structure}(c) appear naturally in Eq.~(7) and are responsible for the hybridization 
between L and L+1-derived bands found in DFT calculations 
\cite{Kosugi09a,Kim11a,deAndres11a,deAndres11b,Kosugi11a}. Fig.~\ref{structure}(d) shows the effect 
of $H_{intra}^{ee}$ on many-electron configurations. 
Here $\Delta_{L,L+1}$ = $0.16|t|$ for phenanthrene for $\epsilon=0$ and is even smaller for realistic 
$\epsilon>0$ \cite{Huang12a}. In contrast, $\Delta_{L,L+1}$ in anthracene is much larger at $0.59|t|$ 
(see Table I).

\begin{figure}[tb]
\includegraphics[width=3.2in]{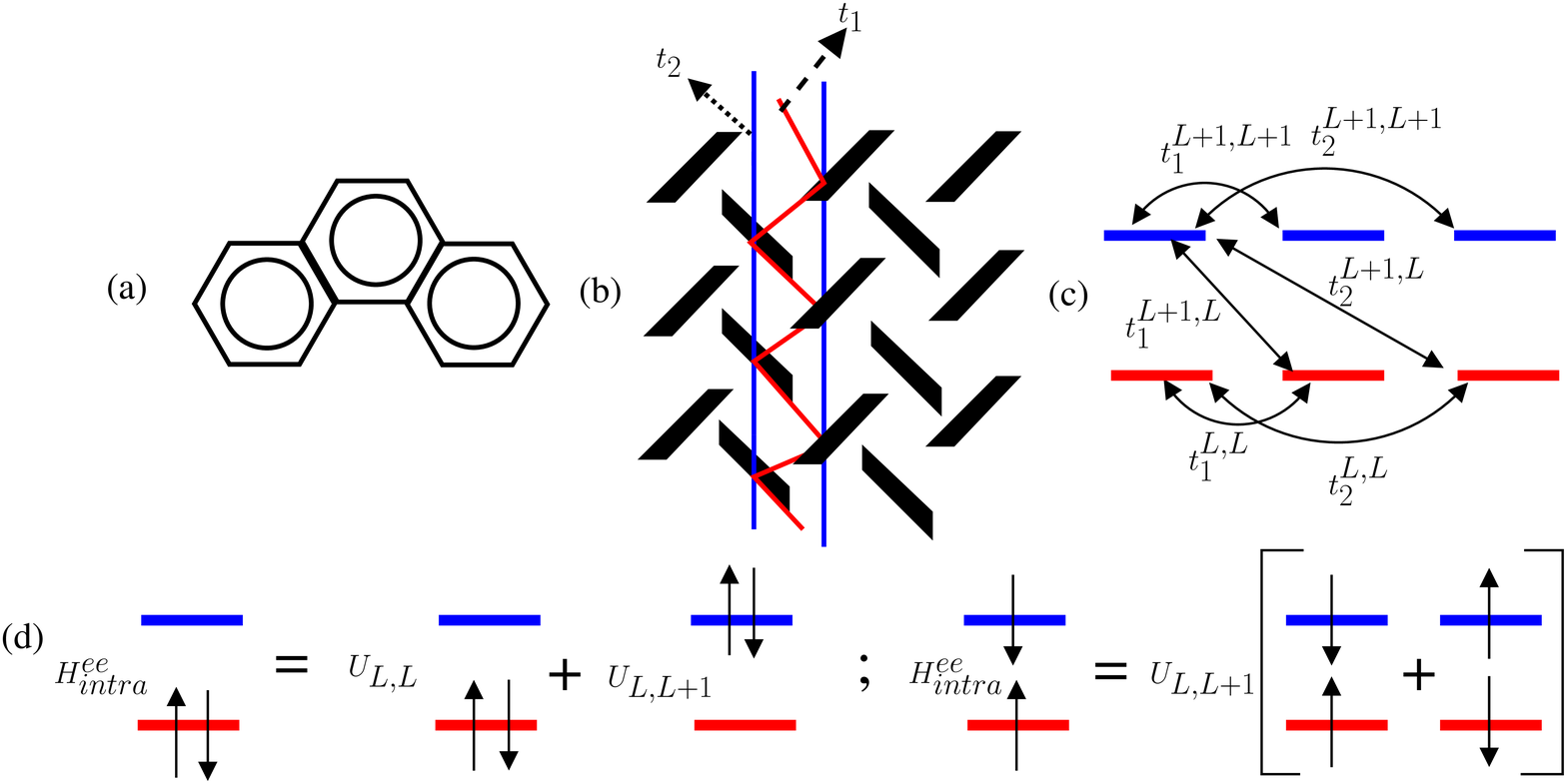}
\caption{(Color online) (a) Molecular structure of phenanthrene. (b) 2D herringbone lattice of doped phenanthrene molecules, with the 1D lattice superimposed on it. Nearest and second neighbor hoppings 
$t_1$ and $t_2$ are indicated. (c) Schematic showing the hopping integrals between L and L+1. (d) 
Effects of Coulomb interactions on different configurations in MO space.}
\label{structure}
\end{figure}

The parameters of the reduced Hamiltonian are the intramolecular $\Delta_{L,L+1}$ and e-e 
interactions, and the multiple intermolecular hopping integrals. The first two are obtained from 
molecular calculations, while we parameterize the intermolecular hoppings based on the existing 
literature on the CTS. Justification for the latter assumption comes from the comparable or even 
larger superconducting T$_c$ in the phenacenes than in the CTS. We then perform exact calculations 
within our reduced Hamiltonian for clusters of phenanthrene ions with charges of $-$1, $-$2 and $-$3, 
and show that there exists a realistic parameter range ($\Delta_{L,L+1}>0.2$ eV) wherein crystals 
consisting of mono and dianions of phenanthrene are one- and two-band Mott-Hubbard semiconductors, 
while crystals of trianions are two-band systems with electron density per localized L ($n_{L}$) and 
L+1 ($n_{L+1}$) close to $\frac{3}{2}$.
Restriction of the parameter range to realistic $\Delta_{L,L+1}>0.2$ eV is necessary, since
for smaller L-L+1 gap the difference between mono and trianions ceases to exist,
and both mono and tri 
anions behave as two-band systems with $\frac{n_{L+1}}{n_{L}}$ $\sim$ $\frac{1}{2}$ and 
$\frac{3}{2}$, respectively.

Exact calculations with $U \neq 0$ are not possible for the 2D lattice of Fig.~\ref{structure}(b). 
Our 2D calculations are therefore for the $U=0$ limit. We choose the standard intramolecular $|t|$ of 
2.4 eV \cite{Baeriswyl92a}, for which $\Delta_{L,L+1}=0.4$ eV at $\epsilon=0$. Although complete band 
structure calculations are possible within Eqs.~(5) and (7) for $U=0$, we perform exact numerical 
calculations within the {\it localized} description because, (a) 
the $U \neq 0$ calculations in 1D reported below are based on the localized basis, and, 
(b) only by considering both zero and non-zero U results 
together, the co-operative effects of intermolecular hopping and Coulomb interactions are fully 
revealed. 
We have calculated numerically exact $n_L$ and $n_{L+1}$, the average L and L+1 occupancies, 
respectively, for a periodic lattice of $20\times20$ molecules ($40\times40$ MOs) for many different 
sets of $t_j^{k_1,k_2}$. In the absence of crystal structure information we did not attempt to 
determine the $t_j^{k_1,k_2}$ from first principles. Rather, our chosen values are comparable to known
hopping integrals in the CTS \cite{Ishiguro}, as already mentioned.
We report computational results for multiple sets of $t_j^{k_1,k_2}$, {\it maximum} values of which 
are, $t_1^{L,L}=0.15$ eV, $t_1^{L+1,L}=0.05$ eV, $t_1^{L+1,L+1}=0.10$ eV; 
$t_2^{L,L}=t_2^{L+1,L+1}=\frac{1}{2}t_1^{L,L}$, $t_2^{L+1,L}=\frac{1}{2}t_1^{L+1,L}$.
For each $(k_1, k_2)$ pair we vary the two $t_j^{k_1,k_2}$ simultaneously, keeping the four other 
hopping integrals fixed at their maximum values.  

\begin{figure}[tb]
\includegraphics[width=3.2in]{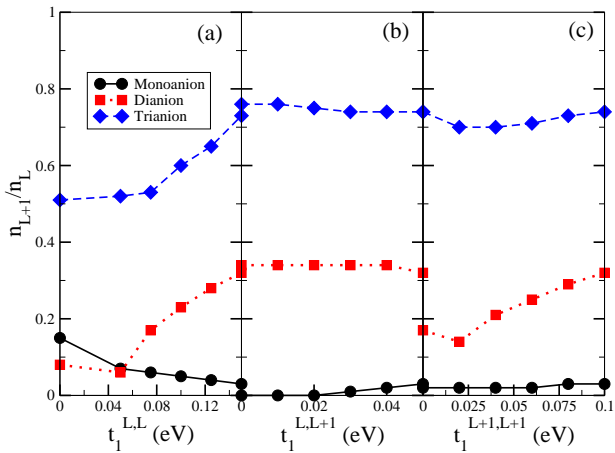}
\caption{(Color online) Ratio of electron populations in L+1 and L, versus different hopping 
integrals, for a periodic 2D lattice of noninteracting electrons, with 20$\times$20 sites (molecules),
and $\Delta_{L,L+1} = 0.4$ eV.}
\label{2D}
\end{figure}

In Figs.~\ref{2D}(a)$-$(c) we report the computational results for all molecular charges $-$1, $-$2 
and $-$3, respectively. 
The monoanions continue to have all electrons in the LUMOs, independent of electron hoppings (except 
in the limit of $t_1^{LL} \to 0$, where nonzero $t_j^{L,L+1}$ promotes a few electrons from L to L+1).
In both the dianion and the trianion, there is a strong enhancement of L+1-population.
%
The different behavior of the monoanions on the one hand, and di and trianions on the other, stem from
the overlap between L- and L+1-derived bands, as demonstrated in the Appendix. As explained there, for all realistic $\Delta$ that is
neither too small nor as large as in anthracene, this difference will persist.

We now demonstrate that e-e interactions further enhance the L+1-populations in the di and trianions,
over and above the bandwidth-driven effect shown in Fig.~\ref{2D}. This is done in 1D. The 1D lattice 
we consider is shown superimposed on the 2D lattice in Fig.~\ref{structure}(a). Band structure 
calculations for K$_3$picene using very similar 1D and 2D lattice motifs give essential features that 
are nearly the same \cite{Casula12a}, justifying our choice of the 1D lattice. Our calculations are 
within the interacting electron Hamiltonian of Eqs. (6) and (7), for periodic 1D clusters of 6 and 
8 molecules (12 and 16 MOs) for all three molecular charges, and 10 molecules (20 MOs) with molecular 
charges $-$1 and $-$3 only. The 10-molecule calculation  with molecular charge $-$2 (20 electrons on 
20 MOs) is beyond current computational capability.   
Numerical results for 8 and 10 molecules are identical for mono and trianions. The calculations are 
done again with parameterized hopping integrals. However, we evaluated the effective e-e interactions 
using Eq.~(6), which gives $U_{L,L}$ = $0.115U$, $U_{L+1,L+1}=0.114U$ and $U_{L,L+1}=0.044U$. For 
realistic atomic $U=8-10$ eV \cite{Baeriswyl92a}, $U_{L,L} \sim U_{L+1,L+1} \approx 1$ eV, close to 
other estimates \cite{Giovannetti11a}. 

\begin{figure}[tb]
\centerline{\resizebox{3.2in}{!}{\includegraphics{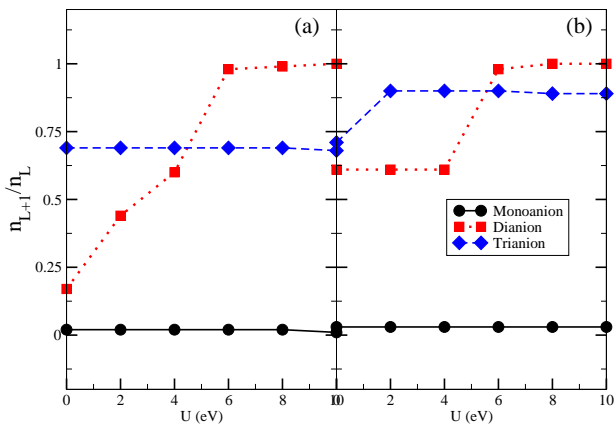}}}
\caption{(Color online) $\frac{n_{L+1}}{n_{L}}$ vs U for interacting electrons on finite clusters of 
10 molecules (mono and trianions), and 8 molecules (dianion), for hopping parameters corresponding 
to the terminal points of Fig.~2. (a) $\Delta_{L,L+1}=$ 0.4 eV, $\epsilon=0$; (b) 
$\Delta_{L,L+1}=0.3$ eV, $\epsilon=0.65$ eV.}
\label{1D}
\end{figure}

Our calculations of the finite clusters at nonzero $U$ are done using the many-body valence bond method of
reference \cite{Chandross99a}. Calculations for all ionic charges were checked by comparing the results of the valence
bond method at $U=0$ against the simple tight binding approach. 
In Figs.~\ref{1D}(a) and (b) we show our numerical results for 10-molecule clusters of mono and 
trianions, and the 8-molecule cluster of dianions, for hopping parameters corresponding to the maximum 
values of Fig.~\ref{2D}. MO populations of the dianion change drastically with $U$, with equally
populated L and L+1 MOs reached at large $U$. 
As is shown in the Appendix, this moderate to large enhancement (depletion) of L+1 (L) electron density 
is driven by $H^{ee}_d$. 
In Fig.~\ref{1D}(a) there is no U-dependence of the MO populations for the monoanion and the 
trianion. However, electrons in the monoanion again occupy only the L, while 
trianion is multi-band. 
Fig.~\ref{1D}(b) differs from \ref{1D}(a) only in having $\epsilon=0.65$ eV (0.27$|t|$), 
which reduces $\Delta_{L,L+1}$ to 0.3 eV. All other parameters are the same. Even with this very 
slight reduction in $\Delta_{L,L+1}$, the consequences for the trianion are dramatic. 
Over and above the bandwidth-driven enhancement in $n_{L+1}$ for $U=0$, further enhancement is seen for nonzero 
$U$, with $n_{L+1}$ approaching $n_L$. As shown in the Appendix, this is a {\it co-operative effect}, driven both
by the bandwidth and  $H^{ee}_d$.

We have found results similar to those in Figs.~\ref{2D} and \ref{1D} for a wide range of hopping 
parameters (see Appendix, section C). The minimal requirement for $n_L \simeq n_{L+1}$ in the di and 
trianions appears to be $t_1^{LL} \sim t_1^{L+1,L+1} \geq$ 0.1 eV, with $t_1^{L,L+1}$ about half 
this value, and second neighbor hoppings close to half the nearest neighbor values. For too small 
$\Delta_{L,L+1}$ the distinction between the mono and the trianion disappears (as noted in the 
Appendix, rigid bond calculations in such cases would be inappropriate). 
We emphasize that the 
parameter range we have considered is very close to those used previously, despite some seeming 
differences. Thus $\Delta_{L,L+1}$ much smaller than ours \cite{Kosugi09a}, as well as nearly the 
same as ours \cite{Giovannetti11a} have been calculated for doped picene. In general, L-L+1 band gaps
obtained from DFT calculations are smaller compared to our $\Delta_{L,L+1}$. 
However, our intermolecular hopping parameters give an overall bandwidth that is close to the upper 
value calculated for doped picene by Subedi and Boeri \cite{Subedi11a}.

Fig.~\ref{1D} suggests that a lattice of phenanthrene monoanions is a one-band Mott-Hubbard 
semiconductor with the L (L+1) band $\frac{1}{2}$-filled (empty). The dianions behave as either 
weakly correlated one-band semiconductor or two-band Mott-Hubbard semiconductor.
In contrast, a lattice of phenanthrene trianions is two-band, with filling close to $\frac{3}{4}$ in 
each band for moderate $\Delta_{L,L+1}$ and $U$ [as in Fig.~\ref{1D}(b)]. It is interesting to note 
that in Reference \onlinecite{Giovannetti11a}, which had presented an effective $\frac{1}{2}$-filled 
band model ($n_{L+1}=1$) for K$_3$picene, the authors had speculated that ``a more accurate treatment 
of electron-electron correlations may revive'' the possibility that K$_3$picene should be described 
as a ``$\frac{3}{4}$-filled two-band system.''

\begin{figure}[tb]
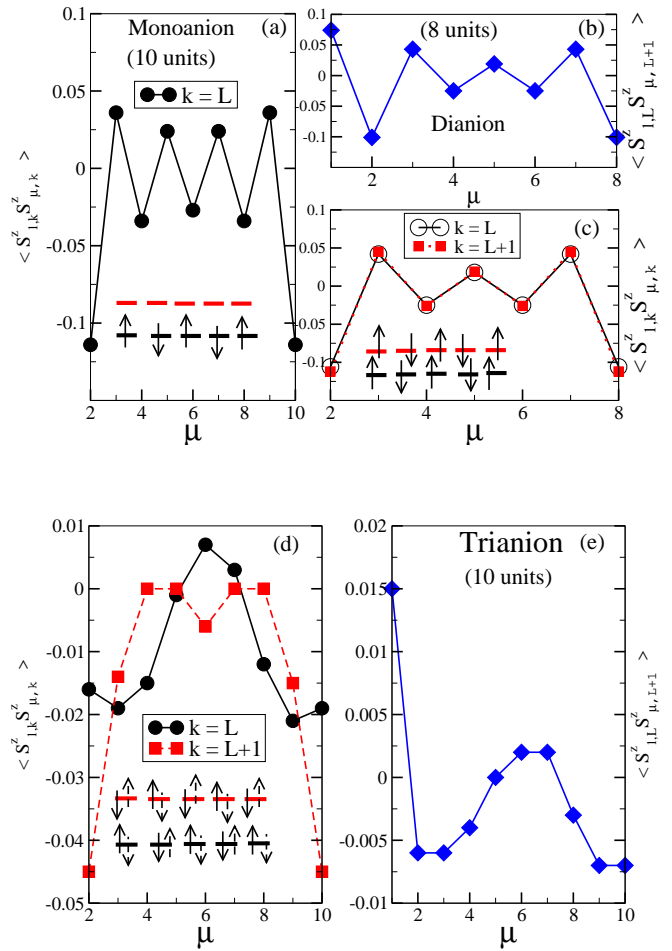

\vspace{0.5cm}
\includegraphics[width=3.4in]{Spin-spin_mono_di.eps} \\
\vspace{0.9cm}
\includegraphics[width=3.4in]{Spin-spin_tri.eps} \\
\caption{Spin-spin correlations ($\Delta_{L,L+1}$ = 0.3 eV, U = 8 eV) for finite clusters of (a) monoanion, 
(b) and (c) dianion, and, (d) and (e) trianion. In all cases, $\mu$ is an index for the molecule. (a) 
L-L correlations in monoanion. L-L+1 and L+1-L+1 correlations are zero and hence not shown. (b) L-L+1 
correlations and (c) L-L and L+1-L+1 correlations in the dianion. (d) shows L-L and L+1-L+1 and 
(e) L-L+1 spin-spin correlations in the trianion. The signs of the correlations in (b) versus (c) 
indicate intra(inter)-molecular ferro(antiferro)magnetic interactions in the dianion. Insets show 
schematic MO occupancies and spins of the electrons occupying them; the ``broken'' arrows in (d) denote
charge $\frac{1}{2}$, giving overall population of $\frac{3}{2}$ per MO in the trianion.}
\label{spin}
\end{figure}

We have calculated spin-spin correlation functions at large $U$ for all three molecular charges. In 
Fig.~\ref{spin} we show our results for the parameters of Fig.~\ref{1D}(b). Both the monoanion and 
the dianion show strong intraband intermolecular antiferromagnetic spin-spin correlations. Notice the 
ferromagnetic intramolecular correlations in the dianion. 
As shown in the Appendix, the intramolecular ferromagnetic correlations and the enhanced $n_{L+1}/n_L$
in the dianion are intimately related.

The spin-spin correlations for the trianion,
in contrast to the other cases, do not indicate any ordering, as is to be expected for the frustrated 
lattice with non-integral occupancies of the MOs. 

The above calculations were repeated for clusters of anthracene ions, and as may be expected from the 
large $\Delta_{L,L+1}$, both mono and trianions are Mott-Hubbard semiconductors now (see Appendix). 
The uniqueness of crystals of trianions of phenanthrene comes from their having 
$n_L \sim n_{L+1} \sim \frac{3}{2}$, which must be a requirement for SC (see next section). This is a 
stronger criterion than having merely multi-band electronic structure, which is certainly reached at 
molecular charges $-2$ and slightly greater values, where however SC is not observed 
\cite{Kubozono11a}. 

\section{Conclusions and Discussions}

We have derived an effective model for crystals of acene and phenacene ions within the 
reduced basis space of L and L+1 orbitals, and performed exact calculations within this basis space 
for clusters of mono, di and trianions, for both atomic $U=0$ and nonzero $U$. We find that 
clusters of trianions of phenanthrene are indeed special, in agreement with what is observed 
experimentally. With realistic $\Delta_{L,L+1}> 0.2$ eV, the electronic population per antibonding MO 
per molecule in this case is $\sim$ $\frac{3}{2}$. 
%
%

Our determination that $n_L \simeq n_{L+1} \simeq \frac{3}{2}$ in the trianions of PAH
suggests that the mechanism of SC in the PAH and the CTS are related.
%
%
Although conducting CTS in general can have a wide range of 
carrier concentration \cite{Keller,Devreese}, SC is limited to compounds with 
carrier concentration of
$\frac{1}{2}$ per molecule, each of which contributes one
active MO. 
Mazumdar and Clay have shown that precisely at carrier concentration $\frac{1}{2}$ and for strong e-e 
and electron-phonon interactions the ground state in both 1D and in frustrated 2D lattices has a 
strong tendency to form a {\it paired-electron crystal} (PEC), in which pairs of charge-rich nearest 
neighbor molecular sites are separated by pairs of charge-poor sites \cite{Clay03a,Li10a,Dayal11a}. 
In 1D this yields a period-4 charge distribution $\cdots1100\cdots$, where `1' and `0' denote the 
charge-rich and charge-poor sites, respectively. The charge distribution is in the form of alternate 
stripes of charge-rich and charge-poor sites in the frustrated 2D lattices, with the charge-rich 
`1-1' sites forming local spin-singlets. 
These authors have suggested that the 
pressure-induced SC in the CTS is a consequence of transition from any of the proximate semiconducting
phases (antiferromagnetic, PEC or spin liquid) to a {\it paired-electron liquid} with mobile local 
spin-singlets, instead of the static singlet pairs that occur in the PEC 
\cite{Mazumdar12a,Mazumdar14a}. 
The local spin-singlets within this approach play the role of the charged bosons in Schafroth's theory of SC \cite{Schafroth55a}.

%
%
Similar local-singlet formation, exactly as in carrier concentration $\frac{1}{2}$, is also expected
for concentration $\frac{3}{2}$, with charge-rich and charge-poor sites simply switched (for 
instance, in 1D, this requires switching from $\cdots1100\cdots$ to $\cdots1122\cdots$). 
Weak deviation of $n_L$ and 
$n_{L+1}$ from exactly $\frac{3}{2}$ [as in Fig.~\ref{1D}(b)] will not affect the pairing, since such 
weak deviations are expected to create positive and negative soliton-like defects within individual 
bands, which will then be paired into interband spin-singlets. The development of a quantitative 
theory of SC based on the proposed scenario clearly requires considerable future work. On the other 
hand, the observation of SC at fixed carrier densities in both PAH and CTS can hardly be coincidences,
and can only be ascribed to e-e interactions. 

Our work has implications for the pressure-induced antiferromagnetic-to-SC transition in A15 
Cs$_3$C$_{60}$ \cite{Ganin08a,Ganin10a}. The ambient pressure antiferromagnetism in this material is explained 
within a theory that incorporates both the Jahn-Teller electron-molecular vibration coupling and the 
Hubbard $U$. 
The Jahn-Teller distortion lifts the three-fold degeneracy of the 
C$_{60}$ LUMOs, which split into a lowest doubly occupied MO, a singly occupied MO at intermediate 
energy, and a vacant MO at the highest energy. Antiferromagnetism is ascribed to the intermolecular spin-spin 
coupling between the unpaired electrons occupying the nondegenerate singly occupied MO \cite{Capone09a}.
Capone {\it et al.} have performed DMFT 
calculations within the $\frac{1}{2}$-filled band Jahn-Teller-Hubbard model, and have proposed that 
pressure-induced increased bandwidth leads to intramolecular pairing of electrons \cite{Capone09a}. 
This conclusion is in apparent disagreement with theoretical works
\cite{Clay08a,Dayal12a,Tocchio09a,Watanabe08a,Gomes13a,Yanagisawa13a} showing the absence of SC within
the $\frac{1}{2}$-filled band (the latter however do not include Jahn-Teller distortion).
An alternate explanation of the antiferromagnetic-to-SC 
transition can be given within our approach, wherein the antiferromagnet at ambient pressure is 
indeed effective $\frac{1}{2}$-filled, but the pressure-induced increase in bandwidth leads to 
equalization of the electron population among the two lowest occupied MOs (as would occur in 
trianions of coronene within our theory, see Appendix). SC in this case would again be related to MO 
populations of $\frac{3}{2}$ in two different MOs, and the pairing is intermolecular. 
We are unaware 
of any experiment that would preclude our proposed mechanism.

To conclude, the trianions of phenacenes are indeed special, with LUMO and LUMO+1 electronic populations of nearly $\frac{3}{2}$
each. The asymmetry between the mono and the trianions is strongly dependent on the {\it cooperative interaction} between the
bandwidth, $\Delta_{L,L+1}$ and $U$. For too large $\Delta_{L,L+1}$ (as in anthracene), as well as for unrealistically small 
$\Delta_{L,L+1}$, this asymmetry vanishes, and molecular charge $-3$ is no longer special. We have proposed that SC is a
consequence of this specific carrier concentration per active MO. Although further work is necessary to actually
demonstrate the transition to the superconducting state, the attractiveness of this proposal
comes from its potential ability to explain SC in all three families of carbon-based superconductors 
within a single theoretical approach.

\section{Acknowledgements}

We are grateful to Professor R. T. Clay for his careful reading of the manuscript.
This work was supported by the U.S. Department of Energy, Office of Science, Basic Energy Sciences, under Award No.
DE-FG02-06ER46315. This research used resources of the NERSC, which is supported by the Office of Science of the U.S. Department of Energy under Contract No. DE-AC02-05CH11231.

\section{Appendix}

\subsection{Mechanism of the valence-dependent electron population exchanges}

In the following we give physical, intuitive explanations of the different behavior of the three ionic 
charges, supported by additional numerical calculations.
Although our discussions are general, we focus largely on the parameters of Fig. 4(b), for the sake of illustration only. 
We discuss band
effects ($U$ = 0) and effect of electron-electron interactions ($U$ $\ne$ 0), separately.

\begin{figure}[tb]
\includegraphics[height=2.4in,width=3.4in]{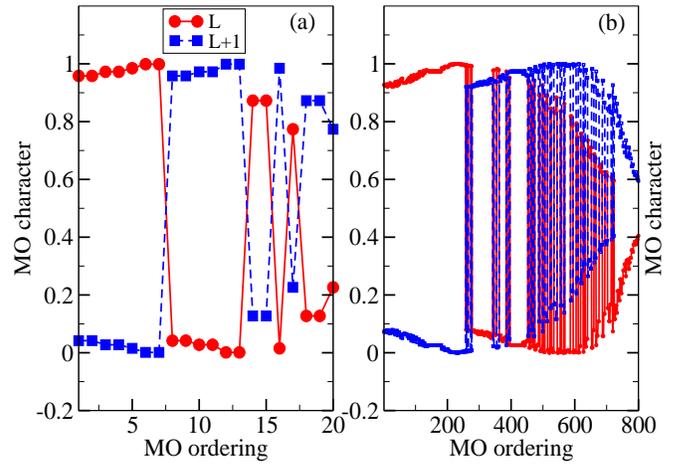} 
\caption{(Color online) 
Normalized L and L+1 characters of the molecular orbitals of a lattice of (a) 10 and, (b) 20$\times$20 
phenanthrene molecules, with $\Delta_{L,L+1}$ = 0.3 eV. The MO ordering is according to increasing
single-particle energies.}
\label{charcter}
\end{figure}

\vskip 0.5pc
{\bf A.1. Band Effects ($U$ = 0):} We have calculated numerically the one-electron eigenspectra within $H^{1e}_{inter}$ (Eq.~7)
for lattices of 10 and 20$\times$20 phenanthrene molecules.
In Figs.~\ref{charcter}(a) and (b) we show the L and L+1 characters of the resulting one-electron eigenstates.
The one-electron energy levels have been numbered in increasing order of energy. For clarity, degenerate levels are given
consecutive indices. The similarity of the 
plots in (a) and (b) indicates that beyond a threshold size (which has already been reached at 10 molecules in 1D), 
band effects in 1D and 2D are same. 
Each eigenstate has both L and L+1 character, the normalized relative weights of which depend on both $\Delta_{L,L+1}$ and
intermolecular hoppings.
As indicated in the figures, for the parameters of Fig.~\ref{1D}(b), the eigenstates are either predominantly
L or predominantly L+1. 
Importantly, while in the lowest energy region the bands are predominantly L-derived, at higher energies there is overlap
between the L- and L+1-derived eigenstates. 
It is easily ascertained that by filling these levels sequentially, only the L-derived bands are occupied
in the monoanions (the lowest 5 MOs accommodate all 10 electrons in the 10-molecule monoanion cluster and the
lowest 200 accommodate all 400 electrons in the 20$\times$20 cluster). Hence $n_{L}$ = 1 here, as calculated in Figs.~3 and 4(b). 
Switchings in the L- and L+1-characters occur at slightly higher energy, even as the bands retain their predominantly L- and
L+1-characters. Sequential occupancy of the MOs in the dianions involve filling of both the 
lowest L- and L+1-derived 
eigenstates, such that $n_{L}$ $<$ 2 and $n_{L+1}$ $>$ 0 (MOs upto the 10th and the 400th are occupied
in the 1D and 2D lattices now). The same band overlap effect (as
seen from Fig.~\ref{charcter}) leads to $n_{L}$ $<$ 2 and $n_{L+1}$ $>$ 1 in the trianion.
{\it We emphasize that we have performed the 1D $U=0$ calculations using both the tight-binding approach
and the valence bond approach we used for $U \neq 0$ \cite{Chandross99a}, and in all cases the calculated $n_{L}$ and
$n_{L+1}$ are identical.} Thus, the band structure effects alone makes the monoanion different from
the di and trianions. 

\begin{figure}[tb]
\includegraphics[height=2.4in,width=3.4in]{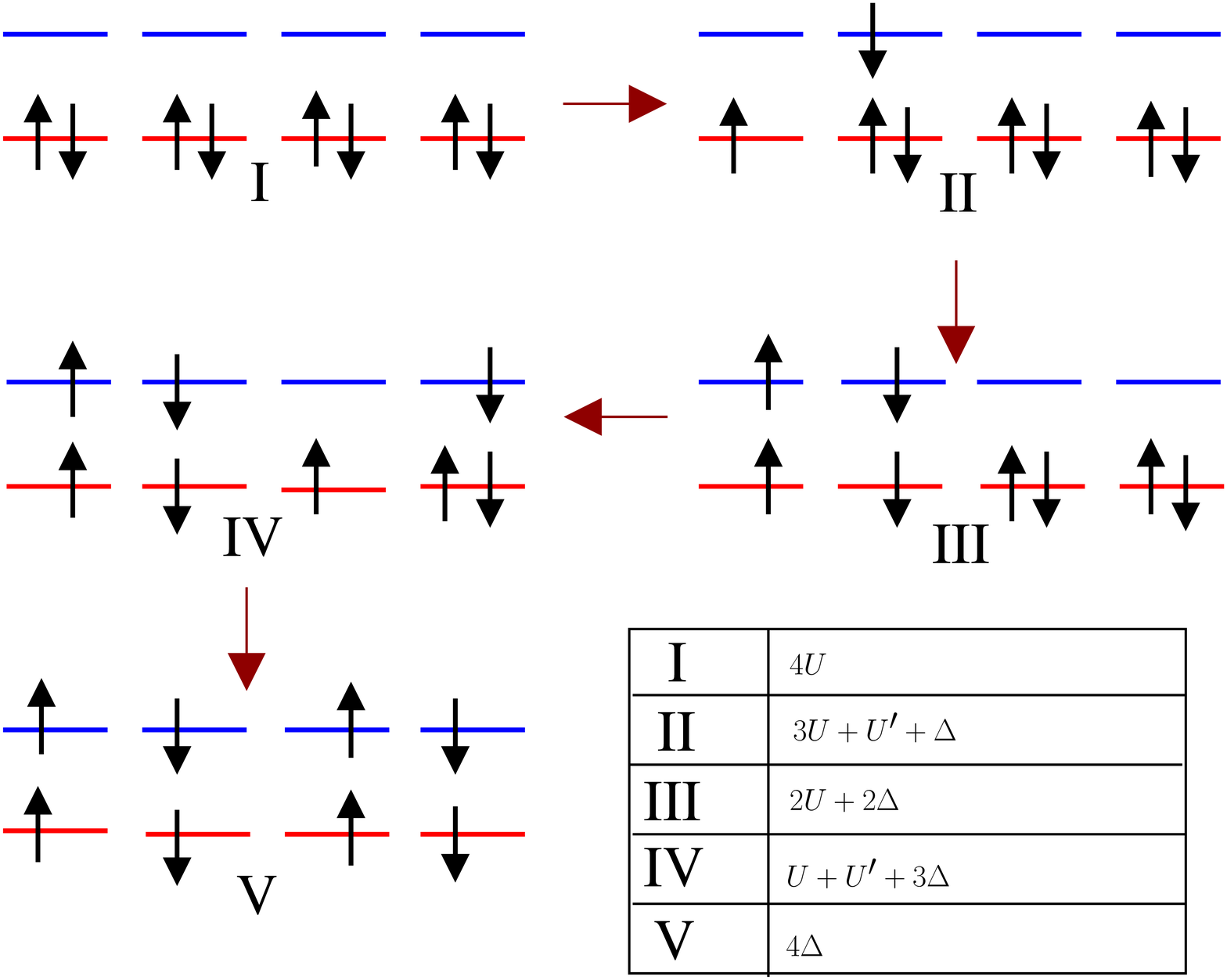}  
\caption{(Color online) 
Schematic illustration of the cooperation between bandwidth and e-e interaction effects leading to
equalization of electron populations in L and L+1 in the dianion (see text). The  
diagonal matrix elements of $H_{L,L+1}$ for configurations $I$ through $V$ are given in the Table, in which 
the L and L+1 orbitals are at single-particle energies 0 and $\Delta$, respectively, where $\Delta$
denotes $\Delta_{L,L+1}$;  
$U$ here
denotes $U_{L,L}~\sim~U_{L+1,L+1}$ and $U^\prime$ denotes $U_{L,L+1}$.
Configurations in which bonding and
antibonding MOs on the same molecule have opposite spins have larger diagonal matrix elements than those with
parallel spins and are not shown.} 
\label{di_path}
\end{figure}

\begin{figure}[tb]
\vspace{0.5cm}
\includegraphics[width=3.0in]{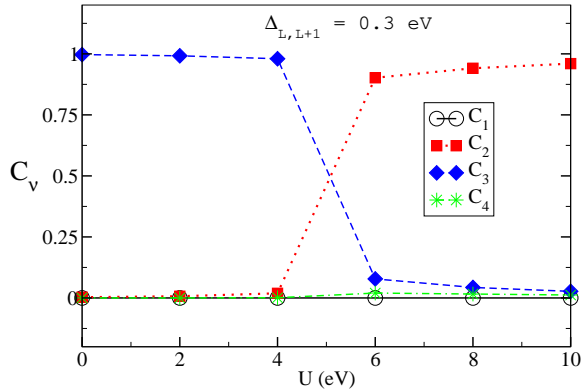} \\
\caption{(Color online)
The sum total of the normalized relative weights of configurations belonging to classes $\nu$=1$-$4 versus $U$ for the
dianion cluster of Fig.~\ref{1D}(b).}
\label{di_wt}
\end{figure}

\vskip 0.5pc
{\bf A.2. Effect of e-e interactions ($U$ $\ne$ 0):} 
For nonzero $U$, our discussions will be in the localized representation which is more appropriate.
\vskip 0.5pc
\noindent{\it Monoanion.} We have seen in the above that finite intermolecular hopping has no effect
on the orbital occupancies in the monoanion for realistic $\Delta_{L,L+1}$. The effect of nonzero $U$
is merely to localize the electrons occupying the L orbitals. The system is a one-band Mott-Hubbard 
semiconductor as seen from spin-spin correlations in Fig.~\ref{spin}(a). For the other ionic charges,
especially for the trianion, we will see that $U$ enhances the bandwidth-induced population exchange 
in a highly cooperative fashion.
\vskip 0.5pc
\noindent{\it Dianion.}  
In Fig.~\ref{di_path}, we have given a schematic path in configuration space between the two extreme 
configurations, with $[n_{L} = 2,n_{L+1} = 0]$ and $[n_{L} = n_{L+1} = 1]$, for the case of four 
molecules, for illustration. Each step in the path constitutes one individual electron hop induced
by $H_{inter}^{1e}$. Similar paths exist in the infinite solid, both in 1D and 2D. The band calculations in
section A.1 indicate that even at $U=0$ the quantum mechanical wavefunction is not described by 
$I$ alone, but has contributions from configurations $II$ and higher. The diagonal matrix elements of 
$H_{L,L+1}$, shown in the Table, decrease from $I$ through $V$ for atomic $U~>$ 2, 
as a consequence of which the ground 
state wavefunction gravitates further towards $V$ with increasing $U$. At 
$U_{LL} \simeq U_{L+1,L+1} \sim \Delta_{L,L+1}$, the transition to $V$ is complete, as seen in 
Figs. 4(a) and (b). 
We have performed analysis of the exact ground state wavefunction in Fig.~\ref{di_wt} for the parameters of 
Fig.~4(b) to illustrate this. We write the ground state wavefunction of the phenanthrene 
dianions as $\Psi=\sum_{\nu,j}A_{\nu,j}|\phi_{\nu,j}\rangle$, where $|\phi_{\nu,j}\rangle$ are the many-electron
configurations and $\sum_{\nu,j}|A_{\nu,j}|^{2}=1$. 
The index $\nu=1-4$ classifies $|\phi_{\nu,j}\rangle$ into four classes 
based on their L and L+1 populations. $\nu=1$ and 2 refer to all configurations with 
{\it integer} $\frac{n_{L+1}}{n_{L}}$ $=$ 0 and 1, respectively, while, $\nu=3$ and 4 correspond to  
configurations with {\it fractional} $\frac{n_{L+1}}{n_{L}}$ $<$ $1$ or $>$ $1$, respectively. The total weight of 
configurations of each class in $\Psi$ is defined as $C_{\nu}=\sum_j|A_{\nu,j}|^{2}$. As seen in Fig.~\ref{di_wt}, the
wavefunction is dominated by configurations with $\nu=3$ at small $U$ including $U=0$, in agreement with the fractional
$\frac{n_{L+1}}{n_{L}}$ seen from the band calculations. A
switching of the ground state occurs at the same $U$ where the jump in $\frac{n_{L+1}}{n_{L}}$ occurs in Fig.~4(b),
indicating that the ground state is dominated by configurations of type $V$.

\begin{figure}[tb]
\vspace{0.5cm}
\includegraphics[height=2.4in,width=3.4in]{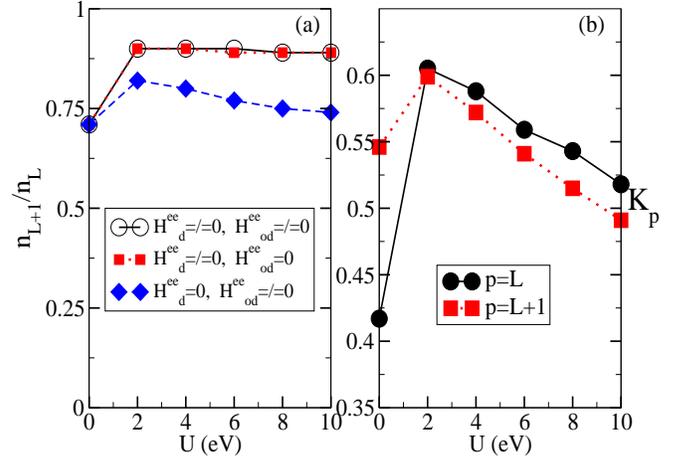} 
\caption{(Color online) (a) $\frac{n_{L+1}}{n_{L}}$ versus $U$ for three different cases; (b) 
Kinetic energies of electrons occupying the L and L+1 orbitals versus U. All parameters correspond to those
of Fig.~4(b).}
\label{tri_keU_effect}
\end{figure}

\vskip 0.5pc
\noindent{\it Trianion.}  
The case of the trianion is more subtle than either the monoanion or the dianion. We have repeated
our exact calculations of Fig.~\ref{1D}(b) by (i) switching off the two-electron hops $H^{ee}_{od}$ between the MOs
but leaving $H^{ee}_d \neq 0$, and, (ii) the
exact opposite: switching off $H^{ee}_d$ but retaining $H^{ee}_{od}$.
We show the results of our calculations in Fig.~\ref{tri_keU_effect}(a). Surprisingly, the 
enhancement of $n_{L+1}$, seen in Fig.~\ref{1D}(b), appears to be driven almost entirely by the
diagonal interactions $H^{ee}_d$. The two-electron hop $H^{ee}_{od}$ has a non-negligible but
weaker effect when $H^{ee}_d=0$, and has no role when $H^{ee}_d \neq 0$. 
We explain these apparently counterintuitive observations below.

\begin{figure*}[tb]
\centering
\includegraphics[width=0.95\textwidth]{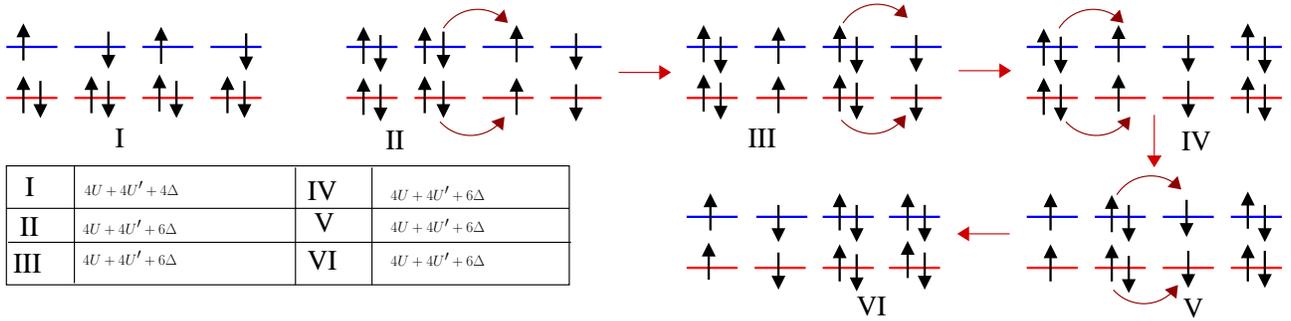}
\caption{(Color online)
The Mott-Hubbard configuration $I$ of the trianion, and five other configurations 
$II-VI$ with $n_L=n_{L+1}$. The Table gives the diagonal matrix elements of $H_{L,L+1}$ for all 
configurations. The curved arrows on each configuration denote single electron hops that conserve 
the number of double occupancies; the flow from $II$ to $VI$ is given by the straight arrows.}
\label{tri_config}
\end{figure*}

Extending our discussion of the dianion in terms of many-electron configurations within the localized
description, we have shown in Fig.~\ref{tri_config}, again for
four molecules, the competing configurations with 
$[n_{L} = 2, n_{L+1} = 1]$ and $[n_{L} = n_{L+1} = 1.5]$. As in Fig.~\ref{di_path}, we have given
the diagonal matrix elements of $H_{L,L+1}$ for each configuration. 
Note that except for configuration $I$, all other
configurations have identical matrix elements for the e-e interaction terms, and are coupled through $H_{inter}^{1e}$.
Our band calculations (Figs.~3 and 6) show that the kinetic energy gain from the electron hopping already drives
$\frac {n_{L+1}}{n_L}$ to greater than 0.5. The important points now are the following:
(i) the one-electron hops involving the L+1-electrons of $I$ necessarily create additional double occupancies 
and are suppressed by $U$.
(ii) In contrast, even as one-electron hops that create additional double occupancies in $II-V$ are also suppressed at finite $U$,
there also exist {\it one-electron hops that conserve the number of double occupancies}. These are suppressed only weakly by $H^{ee}_d$,
and can therefore still couple the
set of configurations with $[n_{L} = n_{L+1} = 1.5]$, as shown in Fig.~\ref{tri_config}.
The number of such double occupancy conserving hops are the largest when $n_{L} = n_{L+1}$, and hence the {\it additional relative
kinetic energy gain} at nonzero $U$ drives the system towards this population ratio. 

We demonstrate the validity of this argument from our exact calculation of the expectation values of the kinetic energies,
\begin{eqnarray}
K_{L} &= \sum_{\mu,\sigma} \biggl( t^{L,L}_{1}\langle a^{\dagger}_{\mu,L,\sigma}a_{\mu+1,L,\sigma} + \text{H.C.}\rangle \nonumber \\ 
      &+ t^{L,L}_{2}\langle a^{\dagger}_{\mu,L,\sigma}a_{\mu+2,L,\sigma} + \text{H.C.}\rangle \biggr),
\\ 
K_{L+1} &= \sum_{\mu,\sigma} \biggl(t^{L+1,L+1}_{1}\langle a^{\dagger}_{\mu,L+1,\sigma}a_{\mu+1,L+1,\sigma} + \text{H.C.}\rangle \nonumber \\
        &+ t^{L+1,L+1}_{2}\langle a^{\dagger}_{\mu,L+1,\sigma}a_{\mu+2,L+1,\sigma} + \text{H.C.}\rangle\biggr). 
\end{eqnarray}
corresponding to the parameters of Fig.~\ref{1D}(b). 
%
We show the results of these 
calculations in Fig.~\ref{tri_keU_effect}(b). There are two significant observations. First, 
while at $U$ = 0, $K_{L+1}$ dominates over $K_{L}$, as is expected for $\frac{n_{L+1}}{n_L}$
only slightly larger than 0.5,  
for all $U \ne$ 0, $K_{L+1} \simeq K_{L}$ in agreement with 
$n_{L}~\simeq~n_{L+1}~\simeq$ 1.5. Secondly, and more importantly, the sum 
$K_{L}+K_{L+1}$ not only shows a jump at small $U$ but is actually larger at all nonzero $U$ values in Fig.~\ref{tri_keU_effect}(b)
than at $U$ = 0, in precise agreement with our argument that the equal probability of $L$ and $L+1$ 
is a consequence of the larger kinetic energy gain, relative to the Mott-Hubbard configuration, at finite $U$ 
(the decrease in the sum with $U$ is primarily due to the suppression of the electron hops that do not conserve the number of
double occupancies). Clearly, this
kinetic energy gain is not sufficient for driving the system to $n_{L}~\simeq~n_{L+1}$ when $\Delta_{L,L+1}$ is very large
(as in anthracene), but as our calculations indicate, this population distribution can occur for the $\Delta_{L,L+1}$ expected
in phenanthrene. 

\begin{figure*}[tb]
\centering
\includegraphics[width=0.60\textwidth]{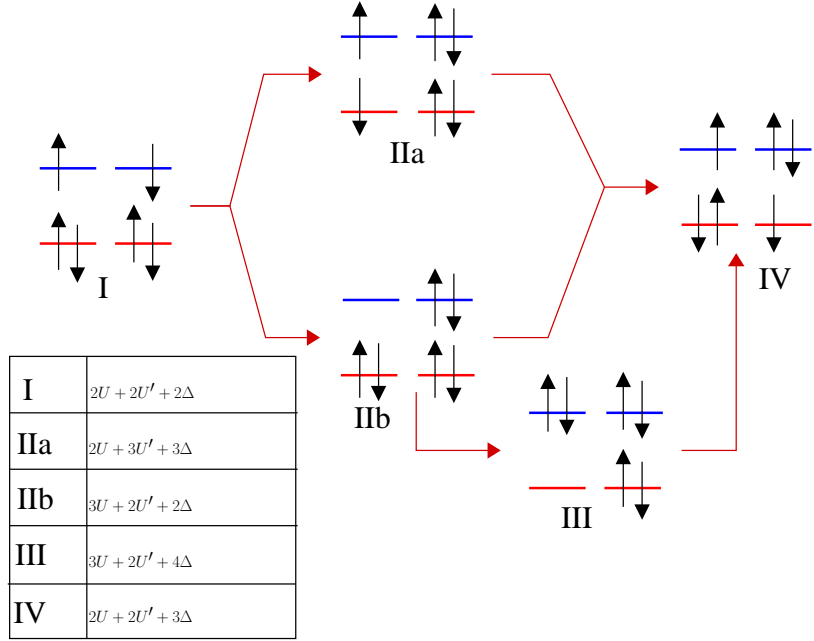}
\caption{(Color Online)
Paths linking the Mott-Hubbard configuration $I$, and the configuration $IV$ with $n_L=n_{L+1}$,
for the case of two molecules (see text). Two of the three paths involve only single electron hops.
$IIb$ to $III$ involves a two-electron hop driven by $H^{ee}_{od}$.}
\label{2e-hop}
\end{figure*}

We now address the role of the $H^{ee}_{od}$. In Fig.~\ref{2e-hop} we have constructed paths connecting the
extreme configurations [$n_L=2$; $n_{L+1}=1$] and [$n_L=n_{L+1}=1.5$], for the case of two molecules.
The attached Table gives the diagonal matrix elements of $H_{L,L+1}$.
Each step in the paths $I \to IIa \to IV$ and $I \to IIb \to IV$ again constitutes a single-electron hop,
and the paths are equally probable at $H^{ee}_d=0$. From the Table, the energies of the intermediate
and final configurations along both paths are larger by $\Delta_{L,L+1}$ at $U=0$. 
The path $I \to IIb \to III \to IV$ involves two-electron
hops and is relevant only for $H^{ee}_{od} \neq 0$. 
Because of this additional channel, for $H^{ee}_d=0$ there
can be enhancement in $\frac{n_{L+1}}{n_L}$ as is seen in Fig.~\ref{tri_keU_effect}(a). 
The enhancement is however weak, for
two reasons: (i) The competing paths not involving two-electron hops are shorter, and (ii) the diagonal energy of 
configuration $III$ is higher by 2$\Delta_{L,L+1}$ at $U=0$. Both (i) and (ii) make the path involving
two-electron hops costly relative to the competing paths.  
{\it Although our illustration is for the case of two
molecules, the same difference between
paths involving and not involving two-electron hops persist in larger systems.} The role
of $H^{ee}_{od}$ in the limit of 
$H^{ee}_d=0$ is both relatively weak and complicated, and determined by the 
detailed values of hopping parameters, $\Delta_{L,L+1}$ and $U$. For $H^{ee}_d \neq 0$ 
there is a new effect that 
makes channels involving two-electron hops even less competing. As seen from Fig.~\ref{2e-hop}, the 
intermediate configuration $III$ is now at even higher diagonal energy because 
of the additional double occupancy. This explains why when both $H^{ee}_d$ and
$H^{ee}_{od}$ are nonzero, $H^{ee}_{od}$ has no effect on $\frac{n_{L+1}}{n_L}$. The weak role of  $H^{ee}_{od}$ is also true for
for the dianion, where we have already shown in Figs.~\ref{di_path} and \ref{di_wt} that $\frac{n_{L+1}}{n_L}$
is given by $H^{ee}_d$ alone.  

\subsection{Application to coronene, picene and anthracene.}

SC in metal-intercalated coronene and picene is also limited to trianions. 
MO occupancies
in these systems are the same as that in phenanthrene: 3 doped electrons occupy the L and L+1
orbitals (see Fig.~\ref{picene1}). However, numerical calculations demonstrating
two nearly filled $\frac{3}{4}$-filled bands in these cases would be far more involved. 
Here we give heuristic physical arguments why results similar to that for phenanthrene should
be expected in these cases also.

\begin{figure}[tb]
\vskip 0.75pc
\includegraphics[height=2.2in,width=3.2in]{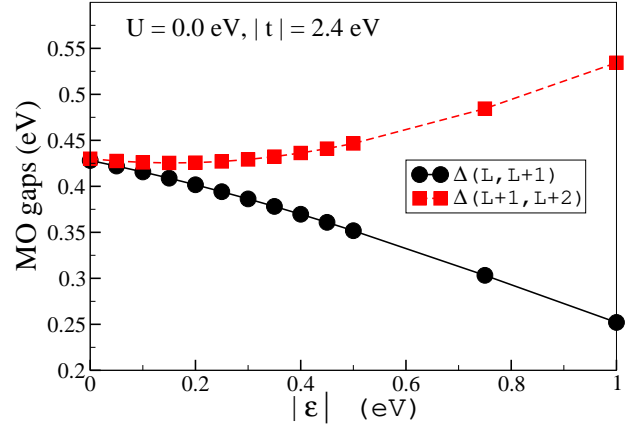} 
\caption{(Color online) 
One-electron energy gaps $\Delta_{L,L+1}$ and $\Delta_{L+1,L+2}$ for picene,
calculated within Eq.~2 for $|t|=2.4$ eV, versus $\epsilon$.}
\label{picene}
\end{figure}

\begin{figure}[tb]
\vskip 0.75pc
\includegraphics[height=2.2in,width=3.2in]{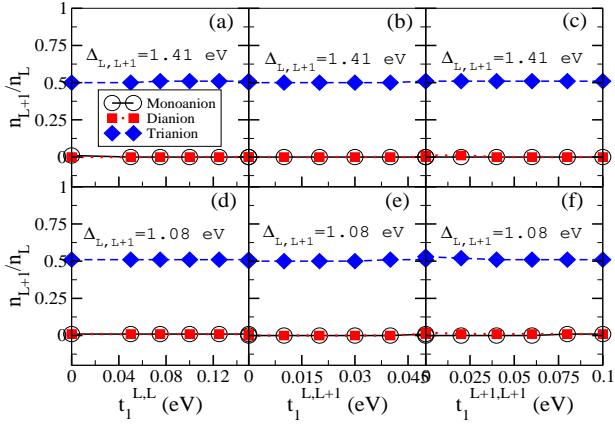}
\caption{(Color online) $\frac{n_{L+1}}{n_{L}}$ versus different hopping integrals at $U=0$,
for 20$\times$20 2D lattice of anthracene ions with parameters of Fig.~\ref{2D}
(a)$-$(c), $\epsilon=0$; (d)$-$(f), $\epsilon=0.65$ eV.}
\label{anthracene}
\end{figure}

\noindent {\it Coronene.} Coronene in the absence of electron-molecular vibration coupling has 
doubly degenerate LUMOs (see Fig.~\ref{picene1})  and thus, the carrier densities per MO are 
expected to be equal, {\it i.e.,} $n_{L_1}=n_{L_2}=\frac{3}{2}$ within a rigid bond model. 
Intramolecular Jahn-Teller distortion lifts this degeneracy, leading to a MO occupied by two 
electrons at lower single-particle energy, and a singly occupied MO at a slightly higher energy. The 
same bandwidth and correlation effects that tend to equalize MO populations in phenanthrene will 
bring back the densities close to $\frac{3}{2}$ here too. It is relevant in 
this context to recall that nonzero Hubbard $U$ reduces the stabilization of doubly occupied MOs due 
to Jahn-Teller distortion, but has no effect on the stabilization energy of singly occupied MOs 
\cite{Dixit84a}. Thus, the Jahn-Teller induced energy gap between the MOs will be larger for the 
monoanion than for the trianion, and an asymmetry between the monoanion and the trianion is expected 
in coronene even without inclusion of intermolecular coupling.
\\
\noindent {\it Picene.} The case of  picene is more interesting, as solution of $H^{1e}_{intra}$
for $\epsilon=0$ gives gaps between LUMO and LUMO+1, and, LUMO+1 and LUMO+2 that are equal, 
$\Delta_{L,L+1}=\Delta_{L+1,L+2}=0.18|t|$ (see Table I). We have suggested in the text that 
$\epsilon \neq 0$ provides a more realistic scenario. As shown in Fig.~\ref{picene}, 
$\Delta_{L,L+1}$ {\it decreases} with $\epsilon$ while $\Delta_{L+1,L+2}$ {\it increases.} For 
$\epsilon \sim 0.65$ eV, as in Fig. \ref{1D}(b), $\Delta_{L,L+1}$ for picene and phenanthrene are 
nearly the same. $\Delta_{L+1,L+2}$ on the other hand is much larger. Even more importantly, the 
significant difference between the monoanion and the trianion in phenanthrene (see Figs.~\ref{2D} and 
~\ref{1D}) indicates that the consequences of interband hopping between the singly occupied LUMO+1 
and vacant LUMO+2 will be small. The very large separation between the LUMO and the LUMO+2 
($\Delta_{L,L+1}+\Delta_{L+1,L+2}$) with increasing $\epsilon$ will on the other hand, preclude 
significant hopping between them. We conclude therefore that the reason why trianions of picene are 
special is the same as in phenanthrene.
\noindent {\it Anthracene}. $\Delta_{L,L+1}$ in anthracene is huge, $0.59|t|$ at $\epsilon=0$
(1.42 eV for $|t|=2.4$ eV, see Table I). Even with nonzero $\epsilon$ this gap stays large. For 
the same lattice structure of Fig.~\ref{structure}(b) we have calculated $n_L$ and $n_{L+1}$ for 
20$\times$20 lattices, as in Fig.~\ref{2D}, for $\epsilon=0$ and 
$\epsilon=0.65$ eV. These results are shown in Fig.~\ref{anthracene}. The absence of SC in doped 
anthracene is understood within our theory. We predict Mott-Hubbard semiconducting behavior for all 
integer valence. It is emphasized that this same scenario is expected in intercalated
pentacene, which has been found to be a Mott-Hubbard semiconductor\cite{Craciun09a}. 
\\
\\

\subsection{Other parameter sets} 

\begin{figure}[tb]
\vspace{1.0cm}
\includegraphics[width=3.2in]{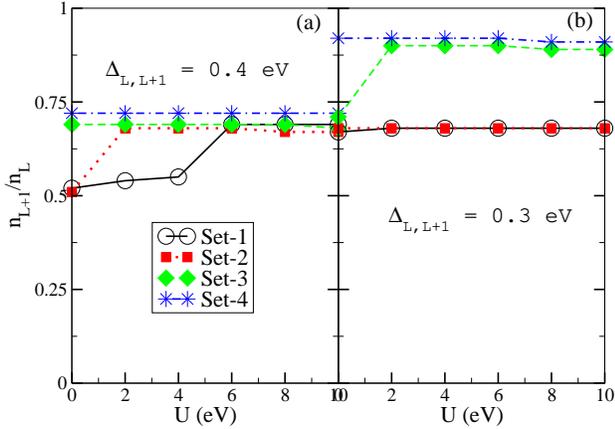}
\caption{(Color online) $\frac{n_{L+1}}{n_{L}}$ vs $U$ for the trianion finite cluster of
Fig.~\ref{structure}, for the hopping parameters of Table I.~ (a) $\epsilon=0$;~(b) $\epsilon=0.65$ eV}
\label{fig-parameters}
\end{figure}

As mentioned in the text, we have performed exact diagonalizations for many different parameter sets, 
for the clusters of Fig.~\ref{1D}. In Table II we have given additional parameter sets beyond those 
in the text for which calculations were done. These include hopping parameters both smaller and 
larger than those of Fig~\ref{1D}. Corresponding to these parameters, in Fig.~\ref{fig-parameters} we 
show change in $\frac{n_{L+1}}{n_{L}}$ with $U$ for phenanthrene trianion, for (a) $\epsilon=0$ 
($\Delta_{L,L+1}=0.4$ eV), and (b) $\epsilon=0.65$ eV, ($\Delta_{L,L+1}=0.3$ eV).

\begin{table}[!htbp]
\caption{Representative parameter sets used in exact diagonalization studies}
\begin{tabular}{|c|c|c|c|c|c|c|c|c|}
\hline
Set & $t_{1}^{L,L}$ & $t_{1}^{L,L+1}$ & $t_{1}^{L+1,L}$ & $t_{1}^{L+1,L+1}$ & $t_{2}^{L,L}$ & $t_{2}^{L,L+1}$ & $t_{2}^{L+1,L}$ & $t_{2}^{L+1,L+1}$ \\
\hline \hline
$1$ & 0.109 & $-$0.059 & 0.059 & 0.143 & 0.05 & 0.0 & 0.0 & 0.05 \\
\hline
$2$ & 0.125 & 0.025 & 0.025 & 0.075 & 0.05 & 0.0 & 0.0 & 0.05 \\
\hline
$3$ & 0.15 & 0.05 & 0.05 & 0.1 & 0.075 & 0.025 & 0.025 & 0.075 \\
\hline
$4$ & 0.175 & 0.075 & 0.075 & 0.125 & 0.1 & 0.05 & 0.05 & 0.1 \\
\hline
\end{tabular}
\end{table}


\end{document}